\documentclass[twocolumn,secnumarabic,amssymb, amsmath, nofootinbib,tightenlines,
nobibnotes, aps, prl,epsfig]{revtex4}
\usepackage{graphicx}
\usepackage{dcolumn}
\usepackage{bm}
\begin{document}
\preprint{APS/123-QED}
\title{The Color Dipole Picture and extracting the ratio of structure functions at small $x$}

\author{B.Rezaei }
\altaffiliation{brezaei@razi.ac.ir}
\author{G.R.Boroun}%
 \email{grboroun@gmail.com; boroun@razi.ac.ir }
\affiliation{ Physics Department, Razi University, Kermanshah
67149, Iran}
\date{\today}
\begin{abstract}
We present a set of formulas to extract the ratio
$F_{L}(x,Q^{2})/F_{2}(x,Q^{2})$ and $R(x,Q^{2})$ from the proton
structure function  parameterized  in the next-to-next-to-leading
order of the perturbative theory at low $x$ values. The behavior
of these ratios  are considered  with respect to the power-law
behavior of the proton structure function. The results are
compared with experimental data and the color dipole model bounds.
These ratios are in good agreement with the HERA data throughout
the fixed value of the invariant mass. The behavior of these
ratios controlled by the nonlinear corrections at low values of
$Q^{2}$. These results and comparison with HERA data demonstrate
that the suggested method for the ratio of structure functions can
be applied in analyses of
the LHeC and FCC-eh projects.\\
\end{abstract}
 \pacs{***}
\maketitle
\subsection{1. Introduction}

Deep inelastic scattering (DIS) can be described in terms of the
imaginary part of forward Compton-scattering amplitude. The
inclusive deep inelastic scattering  measurements are of
importance to understanding the substructure of proton. It is
known that the dominate source for distribution functions is the
gluon density at low values of the Bjorken variable $x$. At low
values of $Q^{2}$ (i.e. $Q^{2} \ll M^{2}_{Z}$)  the contribution
of Z exchange is negligible.  The reduced cross section in terms
of the structure function $F_{2}$ and the longitudinal structure
function $F_{L}$ is defined in the following form:
\begin{eqnarray}
{\sigma_{r}}(x,Q^{2})&=&\frac{d^{2}\sigma}{dxdQ^{2}}.\frac{Q^{4}x}{2{\pi}\alpha^{2}Y_{+}}\nonumber\\
&&=F_{2}(x,Q^{2})[1-\frac{y^{2}}{Y_{+}}\frac{F_{L}(x,Q^{2})}{F_{2}(x,Q^{2})}],
\end{eqnarray}
where $Y_{+}=1+(1-y)^2$ and $\alpha$ is the fine structure
constant. The reduced cross section depends on  the square $s$ of
the center-of-mass energy and the inelasticity variable
$y=Q^{2}/(xs)$ where $Q^{2}$ refers to the photon virtuality. At
high values of $y$  a characteristic bending of the reduced cross
section is observed. It is attributed to the contribution caused
by the longitudinal structure function. The
 structure functions (i.e., $F_{2}$ and  $F_{L}$)
can be written in terms of the $\gamma^{*}p$ total cross section
as
\begin{eqnarray}
F_{2}(x,Q^{2})&=&\frac{Q^{2}}{4\pi^{2}\alpha}[\sigma_{L}^{\gamma^{*}p}(x,Q^{2})+\sigma_{T}^{\gamma^{*}p}(x,Q^{2})],\nonumber\\
\mathrm{and}~~~~~~~~~~~~~~~~~\nonumber\\
F_{L}(x,Q^{2})&=&\frac{Q^{2}}{4\pi^{2}\alpha}\sigma_{L}^{\gamma^{*}p}(x,Q^{2}),
\end{eqnarray}
where the subscript $L$ and  $T$ refer to the transverse and
longitudinal polarization state of the exchanged boson. The ratio
of the longitudinal to transverse cross sections is termed
\begin{eqnarray}
R(x,Q^{2})=\frac{\sigma_{L}(x,Q^{2})}{\sigma_{T}(x,Q^{2})}=\frac{F_{L}}{F_{2}-F_{L}}.
\end{eqnarray}
The longitudinal structure function is directly sensitive to the
gluon density.  Beyond the parton model the $F_{L}$ effects  can
be sizable, hence it can not be longer neglected. Also, the
longitudinal structure function is predominant in cosmic
neutrino-hadron cross section scattering. This behavior for the
longitudinal structure function will  check at the Large Hadron
electron Collider (LHeC) project which runs to beyond a TeV in
center-of-mass energy [1-3]. It is a high-energy lepton-proton and
lepton-nucleus collider based at CERN. The LHeC center-of-mass
energy is $\sqrt{s} \simeq 1.3~ \mathrm{TeV}$ which it is about 30
times the center of mass energy range of ep collisions at hadron
electron ring accelerator (HERA). The kinematics in the
($x,Q^{2}$) plane of the LHeC for electron and positron
neutral-currents reaches $\simeq 1\mathrm{TeV}^{2}$
and $\simeq 10^{-6}$ for $Q^{2}$ and $x$ respectively.\\
The Future Circular Collider (FCC) programme is developing designs
  for a higher luminosity particle collider [3]. In this collider the FCC-eh with $50~ \mathrm{TeV}$ proton
beams colliding with 60 GeV electrons which the center-of-mass
energy is $3.5~ \mathrm{TeV}$. These colliders (i.e., LHeC and
FCC-eh) lead into a region of high parton densities at small
Bjorken $x$. Deep inelastic scattering measurements at FCC-eh and
LHeC will allow the determination of the longitudinal structure
function which its determination was so difficult at HERA.\\
 We
known that HERA have been combined the neutral current (NC)
interaction data for $0.045 \leq Q^{2} \leq
50,000~\mathrm{GeV}^{2}$ and $6 \times 10^{-7}\leq x \leq 0.65$ at
values of the inelasticity $0.005 \leq y \leq 0.95$ [4]. The
highest center-of-mass energy  in deep inelastic scattering of
electrons on protons  was $\sqrt{s}\simeq 320~ \mathrm{GeV}$.
These data that collected from 1992 until 2015 are listed in Table
I. The final measurement of $F_{L}$ at HERA was determined in
Ref.[5]. HERA collected $ep$ collision data with the H1 detector
at a electron beam energy of $27.6~ \mathrm{GeV}$ and proton beam
energies of $920, 575$ and $460~ \mathrm{GeV}$, which allowed a
measurement of structure functions at $x$ values
$6.5{\times}10^{-4}{\leq}x{\leq}0.65$ and $Q^{2}$ values $35~
\mathrm{GeV}^{2} {\leq}Q^{2}{\leq}800 ~\mathrm{GeV}^{2}$. The
variation in inelasticity $y$ was achieved at HERA by comparing
high statistics data at highest energy, $E_{p}=920~ \mathrm{GeV}$,
with about $13 \mathrm{pb}^{-1}$ of data at $460~ \mathrm{GeV}$
and $7
\mathrm{pb}^{-1}$ at $575~ \mathrm{GeV}$.\\
This continues the path of deep inelastic scattering is the best
tool to probe the ratio $F_{L}/F_{2}$ into unknown areas of
physics at very low-$x$ values and new colliders kinematics. In
this region, the gluon distribution has a nonlinear behavior.  The
nonlinear effects are provided by a multiple gluon interaction
which lead to the nonlinear terms in the derivation of the linear
DGLAP evolution equations. Therefore the standard linear DGLAP
evolution equations have been modified by the nonlinear
corrections. Indeed the origin of the shadowing correction, in
pQCD interactions, is primarily considered as the gluon
recombination ($g+g \rightarrow g$) which is simply the inverse
process of gluon splitting ($g \rightarrow g+g$). Gribov, Levin,
Ryskin, Mueller and Qiu (GLR-MQ) [6] performed a detailed study of
these recombination processes. This  widely known as the GLR-MQ
equation and involves the two-gluon distribution per unit area of
the hadron. This equation predicts a saturation behavior of the
gluon distribution at very small $x$ [7-9]. A closer examination
of the small $x$ scattering is resummation powers of
$\alpha_{s}\ln(1/x)$ where leads to the $k_{T}$-factorization form
[10]. In the $k_{T}$-factorization approach the large logarithms
$\ln(1/x)$ are relevant for the unintegrated gluon density in a
nonlinear equation.  Solution of this equation develops a
saturation scale where tame the gluon density
behavior at low values of $x$ and this is an intrinsic characteristic of a dense gluon system.\\
The paper is organized as follows. In sect.2, we give a summary
about the ratio of structure functions based on the color dipole
picture. We will study the ratio $F_{L}/F_{2}$ with respect to the
$F_{2}$ parameterization in section 3. Then we introduce a method
to calculate the ratio $F_{L}/F_{2}$ applying the  effective
exponents behavior. In sect.4 we utilize obtained solution to
calculate the nonlinear behavior of the ratio $F_{L}/F_{2}$ at
hot-spot point. Section 5 contains the  results and discussions.
The behavior of the ratio $F_{L}/F_{2}$ and $R$ are compared with
H1 data at fixed value of invariant mass in this section. Finally,
we give our conclusions in sect.6.\\

\subsection{2. A Short Theoretical Input}

In this section we briefly present the theoretical part of our
analysis. The reader can be refereed to the Refs.[11-16] for more
details. At low $x$ (i.e., $x \ll 1$ where $x \simeq
\frac{Q^{2}}{W^{2}}$ and $W$ refers to the photon-proton
center-of-mass energy), the virtual spacelike photon fluctuates on
the proton  are defined into on-shell quark-antiquark,
$q\overline{q}$, vector state. In this process photon interact
with the proton via coupling of two gluons to the $q\overline{q}$
color dipole. This formalism called the color-dipole picture (CDP)
of low-$x$ DIS [11-13]. The mass of $q\overline{q}$ dipole, in
terms of the transverse momentum $\overrightarrow{k_{\bot}}$ is
realized by
$M^{2}_{q\overline{q}}=\frac{\overrightarrow{k_{\bot}}^{2}}{z(1-z)}$.
Here $\overrightarrow{k_{\bot}}$ define as to the photon direction
and  variable $z$, with $0\leq z \leq 1 $, characterizes the
distribution of the momenta between quark and antiquark. The
lifetime of the $q\overline{q}$ dipole is defined by
$\tau=\frac{W^{2}}{Q^{2}+M^{2}_{q\overline{q}}}\gg
\frac{1}{M_{p}}$. This lifetime is much longer than interaction
time with the target at  small $x$. This condition not only
restricts the kinematical range of the color dipole model to the
region $x\ll 1$ but also saturate the $\gamma^{*}$-proton cross
section
with $x\leq 0.1$ [14-16].\\
In the dipole picture, the total deep inelastic cross-section can
be factorized in the following form
\begin{eqnarray}
\sigma_{L,T}^{\gamma^{*}p}(x,Q^{2})&=&\int dz
d^{2}\mathbf{r}_{\bot}
|\Psi_{\gamma}^{L,T}(\mathbf{r}_{\bot},z(1-z),Q^{2})|^{2}\nonumber\\
&&{\times}\sigma_{q\overline{q}}(\mathbf{r}_{\bot},z(1-z),W^{2}),
\end{eqnarray}
where $\Psi_{\gamma}^{L,T}$ are the appropriate spin averaged
light-cone wave functions of the photon and
$\sigma_{q\overline{q}}(r,z,W^{2})$ is the dipole cross-section
which it related to the imaginary part of the $(q\overline{q})p$
forward scattering amplitude. The square of the photon wave
function describes the probability for the occurrence of a
$(q\overline{q})$ fluctuation.  Here the transverse size related
to the photon
polarization [13-16,17].\\
The ratio of structure functions is expressed in terms of the
longitudinal-to-transverse ratio of the photo absorption cross
sections. This ratio has been defined by
\begin{eqnarray}
\frac{F_{L}(x,Q^{2})}{F_{2}(x,Q^{2})}=\frac{R(x,Q^{2})}{1+R(x,Q^{2})}.
\end{eqnarray}
In Refs.[11-15], authors show that at large $Q^{2}$ (i.e.,
$Q^{2}\gg \Lambda_{sat}^{2}(W^{2})$), the ratio of photo
absorption cross sections is given by the theoretically preferred
value of $\rho$ which
\begin{eqnarray}
R(W^{2},Q^{2})|_{Q^{2}\gg \Lambda^{2}_{sat}(W^{2})
}&=&\frac{1}{2\rho},\nonumber\\
\mathrm{and}~~~\rho_{W}=\rho=\frac{4}{3}.
\end{eqnarray}
The structure function $F_{2}(x,Q^{2})$ has been defined by the
color-dipole cross sections [11], as the leading contribution is
given by the following form
\begin{eqnarray}
F_{2}(x,Q^{2})&=&\frac{Q^{2}}{4\pi^{2}\alpha}\sigma^{\gamma^{*}p}(W^{2},Q^{2})=\frac{Q^{2}}{4\pi^{2}\alpha}(\sigma_{L}^{\gamma^{*}p}(W^{2},Q^{2})\nonumber\\
&&+\sigma_{T}^{\gamma^{*}p}(W^{2},Q^{2})),
\end{eqnarray}
which at large-$Q^{2}$ limit, the equation becomes [12]
\begin{eqnarray}
F_{2}(x,Q^{2})&=&\frac{R_{e^{+}e^{-}}}{36\pi^{2}}(T(W^{2})+\frac{1}{2}L(W^{2})),
\end{eqnarray}
where $R_{e^{+}e^{-}}=10/3$ for even active flavor number.\\
In Refs.[11, 12] as to some assumptions about the sea-quark and
gluon distribution behavior into the kinematic variable $W^{2}$
one obtained that
\begin{eqnarray}
T(W^{2})=\rho L(W^{2}).
\end{eqnarray}
The structure function $F_{2}$ in (8) has been defined by
\begin{eqnarray}
F_{2}(x,Q^{2})&=&\frac{R_{e^{+}e^{-}}}{36\pi^{2}}T(W^{2})(1+\frac{1}{2\rho}).
\end{eqnarray}
Indeed the longitudinal-to-transverse ratio of the photoabsorbtion
cross sections related to  $\rho \equiv \rho(x,Q^{2})$ as
\begin{eqnarray}
R(x,Q^{2})=\frac{\sigma_{L}^{\gamma^{*}p}(W^{2},Q^{2})}{\sigma_{T}^{\gamma^{*}p}(W^{2},Q^{2})}=\frac{1}{2\rho(x,Q^{2})}.
\end{eqnarray}
 The ratio $F_{L}/F_{2}$ is expressed in terms of $\rho$ as we
 have it:
\begin{eqnarray}
\frac{F_{L}(x,Q^{2})}{F_{2}(x,Q^{2})}=\frac{1}{1+2\rho(x,Q^{2})}.
\end{eqnarray}
Factor $2$ originates from the difference between the transverse
and longitudinal photon wave function. Also factor $\rho$  is
associated with different interaction of photons  into
$q\overline{q}$ pairs, $\gamma^{*}_{L,T}\rightarrow
q\overline{q}$. The value of $\rho$ predicted to be $1$ in
Ref.[11] or $\frac{4}{3}$ in Refs.[12, 13]. A similar relation
based on the fit to the experimental data around $\rho=1$ was
found in Ref.[11] in the form
\begin{eqnarray}
\overline{\sigma}_{(q\overline{q})_{T}^{J=1}}(\overrightarrow{l}_{\bot}'^{2},W^{2})=
\overline{\sigma}_{(q\overline{q})_{L}^{J=1}}(\overrightarrow{l}_{\bot}'^{2},W^{2}),
\end{eqnarray}
where they are related to the color-dipole cross sections
concerning the gauge-theory structure as
\begin{eqnarray}
{\sigma}_{(q\overline{q})_p}(\overrightarrow{r}_{\bot},W^{2})&=&\int
d^{2}{l}_{\bot}
\overline{\sigma}_{(q\overline{q})_p}(\overrightarrow{l}_{\bot}^{2},W^{2})\nonumber\\
&&\times(1-e^{-i\overrightarrow{l}_{\bot}\overrightarrow{r}_{\bot}}).
\end{eqnarray}
For the specific value $\rho=1$ (i.e., helicity independent) the
ratio of cross sections $R(W^{2},Q^{2})$ is found to be
$R(W^{2},Q^{2})=0.5$. Therefore the ratio of structure functions
was obtained to be
$F_{L}(W^{2},Q^{2})/F_{2}(W^{2},Q^{2})=1/3=0.333$.\\
In Refs.[12, 13] the relation  between  the $q\overline{q}$-proton
interactions (i.e., Eq.13) was derived by a proportionality factor
$\rho$,
\begin{eqnarray}
\overline{\sigma}_{(q\overline{q})_{T}^{J=1}}(\overrightarrow{l}_{\bot}'^{2},W^{2})=\rho
\overline{\sigma}_{(q\overline{q})_{L}^{J=1}}(\overrightarrow{l}_{\bot}'^{2},W^{2}),
\end{eqnarray}
where
$\rho=\frac{<\overrightarrow{k}^{2}_{\bot}>_{L}}{<\overrightarrow{k}^{2}_{\bot}>_{T}}$.
The average transverse momenta squared with respect to the
longitudinal and transverse photons is defined by the following
forms
\begin{eqnarray}
<\overrightarrow{k}^{2}_{\bot}>_{L,T}=M_{q\overline{q}}^{2}\int_{0}^{1}dz
z(1-z)f_{L,T}(z),
\end{eqnarray}
where
\begin{eqnarray}
f_{L}(z)=6z(1-z)
\end{eqnarray}
and
\begin{eqnarray}
f_{T}(z)=\frac{3}{2}(1-2z(1-z)).
\end{eqnarray}
Then the factor $\rho$ corresponds to
\begin{eqnarray}
\rho=\frac{<\overrightarrow{k}^{2}_{\bot}>_{L}}{<\overrightarrow{k}^{2}_{\bot}>_{T}}=\frac{4}{3},
\end{eqnarray}
which shows $R(W^{2},Q^{2})=3/8=0.375$ and the ratio of structure
functions is proportional to
$F_{L}(W^{2},Q^{2})/F_{2}(W^{2},Q^{2})=3/11=0.273$ [12, 13]. Some
analytical solutions [18, 19] have been shown that in the dipole
model there is a strict bound for the ratio of structure functions
as $\frac{F_{L}(x,Q^{2})}{F_{2}(x,Q^{2})}< 0.27$. In realistic
dipole-proton cross section authors in Ref.[20] have been
discussed that the bound is lower than $0.27$ with the ratio
$\simeq 0.22 $.  In Ref.[21] the ratio $R$ is found at
$R=0.260{\pm}0.050$ which this value is constant at the region
$7.10^{-5}<x<2.10^{-3}$ and $3.5\leq Q^{2} \leq 45~
\mathrm{GeV^{2}}$.  In color dipole model the ratio $R$ leads to
the bound $R\leq 0.372$ [19, 22]. In Ref.[23] ZEUS collaboration
is shown that the overall value of $R$ from both the unconstrained
and constrained fits is $R = 0.105^{+0.055} _{-0.037}$ in a wide
range of $Q^{2}$ values ($5\leq Q^{2} \leq 110~GeV^{2}$). These
results are correspondent to the helicity fluctuations of the
photon on the proton. Our insight into the dynamics of these
fluctuations might determined $\rho(x,Q^{2})$ in comparison with
the measurement data and provides some constraints on the CDP bounds.\\

\subsection{3. Formalism}

In perturbative QCD, the longitudinal structure function in terms
of the coefficient functions at small $x$ is given by
\begin{eqnarray}
x^{-1}F_{L}=<e^{2}>(C_{L,q}{\otimes}q_{s}+C_{L,g}{\otimes}g).
\end{eqnarray}
Here $<e^{k}>$ is the average of the charge $e^{k}$ for the active
quark flavors, $<e^{k}>=n_{f}^{-1}\sum_{i=1}^{n_{f}}e_{i}^{k}$.
The perturbative expansion of the coefficient functions can be
written as [24]
\begin{eqnarray}
C_{L,a}(\alpha_{s},x)=\sum_{n=1}a(t)^{n}c_{L,a}^{n}(x),
\end{eqnarray}
where $n$ is the order in the running coupling constant. The
running coupling constant in the high-loop corrections of the
above  equation is expressed entirely  thorough the variable
$a(t)$, as $a(t)=\frac{\alpha_{s}}{4\pi}$. The explicit expression
for the coefficient functions in LO up to NNLO are presented in
Refs.[25, 26].\\
At small $x$  the linear DGLAP $Q^{2}$ evolution equations [27]
related to the BFKL [28] type of law. In this theoretical
framework, the gluon density is expressed in terms of the
structure functions, $F_{2}$ and $F_{L}$, which it is a test
higher order QCD at small $x$. In this respect, several equations
have been proposed to define the gluon density by the $F_{2}$
scaling violation. In Refs.[29, 30] a similar relation based on
the $F_{2}$ and derivative of $F_{2}$ related to the pomeron
behavior was suggested with the result
\begin{eqnarray} G(x,Q^{2})&=&
\frac{1}{\Theta_{qg}(x,Q^{2})}[\frac{{\partial}F_{2}(x,Q^{2})}{{\partial}{\ln}Q^{2}}\nonumber\\
&&-\Phi_{qq}(x,Q^{2})F_{2}(x,Q^{2})].
\end{eqnarray}
The kernels for the quark and gluon sectors (denoted by $\Phi$ and
$\Theta$) presented  by the following forms
\begin{eqnarray}
\Theta_{qg}(x,Q^{2})&=&P_{qg}(x,\alpha_{s})\otimes
x^{\lambda_{g}},\nonumber\\
\Phi_{qq}(x,Q^{2})&=&P_{qq}(x,\alpha_{s})\otimes x^{\lambda_{s}}.
\end{eqnarray}
The one, two- and three-loop splitting functions (LO, NLO and
NNLO) for the parton distributions have been shown in Ref.[31].
The exponents $\lambda_{s}$ and $\lambda_{g}$ are defined by the
derivatives of the distribution functions by the following forms
as
 $ \lambda_{s}={\partial \ln F_{2}^{s}(x,Q^{2})}/{\partial
\ln(1/x)}$ and $ \lambda_{g}={\partial \ln G(x,Q^{2})}/{\partial
\ln(1/x)}$. The running coupling constant $\alpha_{s}$ has the
following form in NNLO analysis
\begin{eqnarray}
\alpha_{s}^{\rm
NNLO}&=&\frac{4\pi}{\beta_{0}t}[1-\frac{\beta_{1}{\ln}t}{\beta_{0}^{2}t}+\frac{1}{(\beta_{0}t)^{2}}
[(\frac{\beta_{1}}{\beta_{0}})^{2}\nonumber\\
&&(\ln^{2}t-{\ln}t+1)+\frac{\beta_{2}}{\beta_{0}}]],
\end{eqnarray}
where $\beta_{0}=\frac{1}{3}(33-2n_{f})$,
$\beta_{1}=102-\frac{38}{3}n_{f}$ and
$\beta_{2}=\frac{2857}{6}-\frac{6673}{18}n_{f}+\frac{325}{54}n_{f}^{2}$.
The variable $t$ is defined as
$t={\ln}(\frac{Q^{2}}{\Lambda^{2}})$ and $\Lambda$ is the QCD
cut- off parameter for each heavy quark mass threshold as we take the $n_{f}=4$ for $m_{c}^{2}<\mu^{2}<m^{2}_{b}$.\\
In what follows it is convenient to use a similar method was found
in Ref.[32] for the longitudinal structure function  as we have
it:
\begin{eqnarray}
F_{L}(x,Q^{2})&=&F_{2}(x,Q^{2})I_{L,q}(x,Q^{2})\nonumber\\
&&+G(x,Q^{2})I_{L,g}(x,Q^{2}),
\end{eqnarray}
where the analytical results for the compact form of the kernels
 are given in Appendix A.\\
With respect to Eq.22, one can rewrite the longitudinal structure
function concerning  the proton structure function
$F_{2}(x,Q^{2})$ and its derivative $\partial
F_{2}(x,Q^{2})/\partial{\ln}Q^{2}$, as we will have
\begin{eqnarray}
F_{L}(x,Q^{2})&=&\frac{I_{L,g}(x,Q^{2})}{\Theta_{qg}(x,Q^{2})}\frac{{\partial}F_{2}(x,Q^{2})}{{\partial}{\ln}Q^{2}}+\{I_{L,q}(x,Q^{2})\nonumber\\
&&-\Phi_{qq}(x,Q^{2})\frac{I_{L,g}(x,Q^{2})}{\Theta_{qg}(x,Q^{2})}
\}F_{2}(x,Q^{2}).
\end{eqnarray}
Therefore the ratio $F_{L}/F_{2}$ takes the form
\begin{eqnarray}
\frac{F_{L}(x,Q^{2})}{F_{2}(x,Q^{2})}=\eta(x,Q^{2})+\zeta(x,Q^{2})\frac{{\partial}{\ln}F_{2}(x,Q^{2})}{{\partial}{\ln}Q^{2}},
\end{eqnarray}
where
 \begin{eqnarray}
 \eta(x,Q^{2})&=&I_{L,q}(x,Q^{2})-\Phi_{qq}(x,Q^{2})\frac{I_{L,g}(x,Q^{2})}{\Theta_{qg}(x,Q^{2})},\nonumber\\
 \zeta(x,Q^{2})&=&\frac{I_{L,g}(x,Q^{2})}{\Theta_{qg}(x,Q^{2})}.
 \end{eqnarray}
 By means of these equations we have extracted the ratio of
 structure functions from the parameterization of $F_{2}$, using the slopes ${\partial}{\ln}F_{2}/{\partial}{\ln}Q^{2}$ proposed in
 Ref.[33].\\

\textbf{3.1.} \textit{The ratio of structure functions using the
$F_{2}$ parameterization}\\

Authors of Ref.[33] suggested a new parametrization which
describes fairly good the available experimental data on the
proton structure function in an agreement with the Froissart bound
behavior. Also a theoretical analysis investigated the behavior of
the longitudinal structure function at small $x$ by employing this
behavior is studied in Ref.[34]. The explicit expression for the
$F_{2}$ parametrization [33] is given by the following form
\begin{eqnarray}
F^{\gamma p}_{ 2}(x,Q^{2})& =& D(Q^{2})(1-
x)^{n}\sum_{m=0}^{2}A_{m}(Q^{2})L^{m},
\end{eqnarray}
where
\begin{eqnarray}
A_{0}(Q^{2})& =& a_{00} + a_{01}
{\ln}(1+\frac{Q^{2}}{\mu^{2}}),\nonumber\\
 A_{1}(Q^{2})& =& a_{10} + a_{11} {\ln}(1+\frac{Q^{2}}{\mu^{2}}) + a_{12}{\ln}^{2}(1+\frac{Q^{2}}{\mu^{2}})
 ,\nonumber\\
A_{2}(Q^{2})& =& a_{20} + a_{21} {\ln}(1+\frac{Q^{2}}{\mu^{2}}) +
a_{22}{\ln}^{2}(1+\frac{Q^{2}}{\mu^{2}})
 ,\nonumber\\
D(Q^{2})& =& \frac{Q^{2}(Q^{2}+\lambda M^{2})}{(Q^{2}+M^{2})^2},\nonumber\\
L^{m}&=&\ln^{m}(\frac{1}{x}\frac{Q^{2}}{Q^{2}+\mu^{2}}).
\end{eqnarray}
Here $M$ and $\mu^{2}$ are the effective mass  a scale factor
respectively. The additional parameters with their statistical
errors are given in Table II. Equation (29) obtained from a
combined fit of the H1 and ZEUS collaborations data [35] in a
range of the kinematical variables $x$ and $Q^{2}$( $x<0.01$ and
$0.15<
Q^{2}<3000~\mathrm{GeV} ^{2}$).\\
 Therefore with respect to the
parameterization of $F_{2}$ the final result for the ratio
$F_{L}/F_{2}$ is
\begin{eqnarray}
\frac{F_{L}(x,Q^{2})}{F_{2}(x,Q^{2})}&=&\eta(x,Q^{2})+\zeta(x,Q^{2})\{\frac{{\partial}{\ln}D(Q^{2})}{{\partial}{\ln}Q^{2}}\nonumber\\
&&+\frac{{\partial}{\ln}(\sum_{m=0}^{2}A_{m}(Q^{2})L^{m})}{{\partial}{\ln}Q^{2}}\}.
\end{eqnarray}
At last, the ratio of the longitudinal to transverse cross
sections is expressed in terms of the $F_{2}$ parameterization as
we have it:
\begin{eqnarray}
R(x,Q^{2})=Eq.(31)/(1-Eq.(31)).
\end{eqnarray}\\

\textbf{3.2.} \textit{The ratio of structure functions using the
exponents behavior}\\

At small $x$ and sufficiently large $Q^{2}$, the deep inelastic
scattering is recognized as elastic diffractive forward scattering
of the photon fluctuations,$(q\overline{q})^{J=1} _{L,T}$, on the
proton [11-15]. The proton structure function in CDP is given by
the single variable $W^{2}$ as
\begin{eqnarray}
F_{2}(x,Q^{2})=F_{2}(W^{2}=\frac{Q^{2}}{x}).
\end{eqnarray}
A power-law for the $W$-dependence of $F_{2}(W^{2})$ can be
exploited in Eq.(33) as
\begin{eqnarray}
F_{2}(W^{2})\sim
(W^{2})^{\lambda_{s}}=(\frac{Q^{2}}{x})^{\lambda_{s}},
\end{eqnarray}
where the exponent $\lambda_{s}$ is defined by
\begin{eqnarray}
\lambda_{s}&=&\frac{\partial \ln F_{2}(W^{2})}{\partial \ln
W^{2}}.
\end{eqnarray}
Substitution of (35) into (27) we get
\begin{eqnarray}
\frac{F_{L}(W^{2})}{F_{2}(W^{2})}=\eta(W^{2})+\zeta(W^{2})\lambda_{s}|_{x}.
\end{eqnarray}
Equivalently the ratio of structure functions, in terms of
$\rho_{w}$, is obtained by the following form
\begin{eqnarray}
\frac{1}{1+2\rho_{w}}=\eta(W^{2})+\zeta(W^{2})\lambda_{s},
\end{eqnarray}
where
\begin{eqnarray}
\rho_{w}=\frac{1-\eta(W^{2})-\zeta(W^{2})\lambda_{s}}{2(\eta(W^{2})+\zeta(W^{2})\lambda_{s})},
\end{eqnarray}
or
\begin{eqnarray}
R_{w}=\frac{\eta(W^{2})+\zeta(W^{2})\lambda_{s}}{1-\eta(W^{2})-\zeta(W^{2})\lambda_{s}}.
\end{eqnarray}
According to the Regge phenomenology,  the density functions can
be controlled by pomeron exchange at low $x$ since these behaviors
for the singlet and gluon distributions are correspondent to the
BFKL pomeron. The exponent for the gluon distribution is
comparable with the so-called hard pomeron intercept. This
behavior is defined with a value  of $\lambda_{g}{\simeq}0.424$
which it is the hard pomeron part of Regge phenomenology [36]. We
choice that the singlet exponent value is consistent with the
experimental data and CDP bound if one definite value $0.27 \leq
\lambda_{s} \leq 0.33$ in a wide range of $Q^{2}$ values.\\
The strong rise into the $k_{T}$ factorization formula is also
true for the singlet structure function.  This behavior comes from
resummation of large powers of $\alpha_{s} \ln1/x$ where its
achieved by the use of the $k_{T}$ factorization formalism. The
small-$x$ resummation requires an all-order class of subleading
corrections in order to lead to stable results [37]. However the
singlet effective pomeron is $Q^{2}$-dependent when structure
functions fitted to the experimental data at low values of $x$.
Here, we take into account the effects of kinematics which lead to
a shift from the pomeron exponent to the
effective exponent for singlet structure function.\\
 To better illustrate our calculations
at all $Q^{2}$ values, we used the singlet effective exponent in
the form of $\lambda^{s}(Q^{2})$. The singlet effective exponent
is presented based on the H1 and H1-ZEUS combined data for the
proton structure functions at $4 < Q^{2} < 200~GeV^{2}$. In
Ref.[38], an eyeball fit was given by the following form
\begin{eqnarray}
\lambda^{s}_{eff}(Q^{2})=0.13+0.1(\frac{Q^{2}}{10})^{0.35}.
\end{eqnarray}
Also authors in Ref.[39]  have derived  the phenomenological
exponent of singlet density for combined HERA $e^{+}p$ DIS data
[40] within the saturation model  where $\lambda^{s}_{phn}(Q^{2})$
is assumed to be of the form
\begin{eqnarray}
\lambda^{s}_{phn}(Q^{2})=0.329+0.1 \log(\frac{Q^{2}}{90}).
\end{eqnarray}

\subsection{4. Nonlinear correction to the ratio $F_{L}/F_{2}$}

The behavior of the singlet density will be checked at new
colliders ( LHeC and FCC-eh) which runs to beyond a TeV in
center-of-mass energy. Clearly, there would be an increase in the
precision of parton density and in the low $x$ kinematic region in
these colliders. So one should consider the low- $x$ behavior of
the singlet distribution using the nonlinear GLR-MQ evolution
equation. The shadowing correction to the evolution of the singlet
quark distribution can be written as [41]
\begin{eqnarray}
\frac{{\partial}xq(x,Q^{2})}{{\partial}{\ln}Q^{2}}&=&\frac{{\partial}xq(x,Q^{2})}{{\partial}{\ln}Q^{2}}|_{DGLAP}\nonumber\\
&&-\frac{27\alpha_{s}^{2}}{160R^{2}Q^{2}} [xg(x,Q^{2})]^{2}.
\end{eqnarray}
Eq.(42) can be rewrite in a convenient form as
\begin{eqnarray}
\frac{{\partial}F_{2}(x,Q^{2})}{{\partial}{\ln}Q^{2}}=\frac{{\partial}F_{2}(x,Q^{2})}{{\partial}lnQ^{2}}|_{DGLAP}-
\frac{5}{18}\frac{27\alpha_{s}^{2}}{160R^{2}Q^{2}}\nonumber\\
{\times}[xg(x,Q^{2})]^{2}.
\end{eqnarray}
The first term is the standard DGLAP evolution equation and the
value of $R$ is the correlation radius between two interacting
gluons. It will be  of the order of the proton radius
$(R\simeq5\hspace{0.1cm} GeV^{-1})$ if the gluons are distributed
through the whole of proton, or much smaller
$(R\simeq2\hspace{0.1cm} GeV^{-1})$ if gluons are concentrated in
hot- spot within the proton.\\
Combining Eqs. (22) and (43), one can consider the nonlinear
correction to the gluon distribution function as we have
\begin{eqnarray}
G(x,Q^{2})&=&
\frac{1}{\Theta_{qg}(x,Q^{2})}[\frac{{\partial}F_{2}(x,Q^{2})}{{\partial}{\ln}Q^{2}}
-\frac{5}{18}\frac{27\alpha_{s}^{2}}{160R^{2}Q^{2}}{\times}\nonumber\\
&&G^{2}(x,Q^{2})-\Phi_{qq}(x,Q^{2})F_{2}(x,Q^{2})].\nonumber\\
\end{eqnarray}
Eq.(44) can be rewritten in the following form:
\begin{eqnarray}
G(x,Q^{2})+\frac{1}{\Theta_{qg}(x,Q^{2})}\frac{5}{18}\frac{27\alpha_{s}^{2}}{160R^{2}Q^{2}}G^{2}(x,Q^{2})=\nonumber\\
\frac{1}{\Theta_{qg}(x,Q^{2})}[\frac{{\partial}F_{2}(x,Q^{2})}{{\partial}{\ln}Q^{2}}-\Phi_{qq}(x,Q^{2})F_{2}(x,Q^{2})].
\end{eqnarray}
Equation (45) is a second-order equation related to the gluon
distribution function which can be solved as
\begin{eqnarray}
\mathcal{G}(x,Q^{2})&=&G(x,Q^{2})[1-\frac{\mathcal{A}(x,Q^{2})}{\Theta_{qg}(x,Q^{2})}G(x,Q^{2})\nonumber\\
&&+2(\frac{\mathcal{A}(x,Q^{2})}{\Theta_{qg}(x,Q^{2})}G(x,Q^{2}))^{2}\nonumber\\
&&-5(\frac{\mathcal{A}(x,Q^{2})}{\Theta_{qg}(x,Q^{2})}G(x,Q^{2}))^{3}+....]\nonumber\\
&&=G(x,Q^{2})[1-\mathcal{N}+2\mathcal{N}^{2}-5\mathcal{N}^{3}+....]\nonumber\\
&&=G(x,Q^{2})[\mathcal{NLC}],
\end{eqnarray}
where
\begin{eqnarray}
\mathcal{N}=\frac{\mathcal{A}(x,Q^{2})}{\Theta_{qg}(x,Q^{2})}G(x,Q^{2}),\nonumber
\end{eqnarray}
and
\begin{eqnarray}
\mathcal{A}(x,Q^{2})=\frac{5}{18}\frac{27\alpha_{s}^{2}}{160R^{2}Q^{2}}.\nonumber
\end{eqnarray}
 Therefore the nonlinear correction ($\mathcal{NLC}$) to the
 ratio $F_{L}/F_{2}$
is obtained by the following form as
\begin{eqnarray}
\frac{F_{L}(x,Q^{2})}{F_{2}(x,Q^{2})}|_{Nonlinear}=I_{L,q}(x,Q^{2})+\frac{I_{L,g}(x,Q^{2})}{\Theta_{qg}(x,Q^{2})}\nonumber\\
{\times}\{(\frac{{\partial}{\ln}D(Q^{2})}{{\partial}{\ln}Q^{2}}
+\frac{{\partial}{\ln}(\sum_{m=0}^{2}A_{m}(Q^{2})L^{m})}{{\partial}{\ln}Q^{2}})\nonumber\\
-\Phi_{qq}(x,Q^{2})\}[\mathcal{NLC}].~~~~~~~~~~~~~~~~~~~~~~~~~~~~
\end{eqnarray}
Indeed the nonlinear correction to the ratio
$\frac{\sigma_{L}}{\sigma_{T}}$ is defined
\begin{eqnarray}
R(x,Q^{2})|_{Nonlinear}=Eq.(47)/(1-Eq.(47)).
\end{eqnarray}

\subsection{5. Results and Discussions}

In this paper, we obtained the ratio of structure functions based
on the $F_{2}$ parameterization and  singlet exponent behavior at
NNLO analysis. The behavior of ratios compared  with HERA data and
also with CDP bounds. We use the $F_{2}$ parameterized [33] where
fitted to the combined H1 and ZEUS inclusive DIS data [35] in a
range of the kinematical variables $x<0.01$ and
$0.15~\mathrm{GeV^{2}}<Q^{2}<3000~\mathrm{GeV^{2}}$. The coupling
constant defined  via the $n_{f}=4$ definition of $\Lambda_{QCD}$
for the ZEUS data [35] and the MRST set of partons [42]. The
values of $\Lambda_{QCD}$ at LO up to NNLO
 are displayed in Table III respectively. The values of $\lambda_{s}\simeq 0.33$ and
$\lambda_{g}\simeq 0.43$ [43-46] are used within the range of
$Q^{2}$ under study. The predictions for  the ratio of structure
functions have been depicted at fixed value of the invariant mass
W (i.e. $W=230~ \mathrm{GeV}$),
and compared in the  HERA kinematic range [5] at low values of $x$.\\
In Fig.1, we show the prediction of Eq.(31) for the ratio
$F_{L}/F_{2}$ and compare this ratio with the H1 data [5] as
accompanied with total errors. As can be seen in this figure, the
depletion and enhancement in this ratio reflect the experimental
data and it is comparable with the H1 data in the interval
$1~\mathrm{GeV^{2}}<Q^{2}<500~\mathrm{GeV^{2}}$. The error bares
are in accordance with the statistical errors of the $F_{2}$
parameterization as presented in Table II. Also a detailed
comparison with the CDP bounds has  been shown  in this figure
(i.e., Fig.1). As can be seen, the values of the ratio
$F_{L}/F_{2}$  are in good agreement with the CDP bounds at $1 <
Q^{2}\leq 20~\mathrm{GeV}^{2} $ at fixed value of the invariant
mass.\\
The  ratio $R$ is expected to vanish at large $Q^{2}$ and moderate
$x$ values in the naive parton model, but it is nonzero at low
values of $x$. It dues to the fact that partons can carry
transverse momentum [47]. In Fig.(2) we present the ratio $R$
related to Eq.(32) in comparison with the
 H1 data  using the $F_{2}$ parameterization. As can be seen in this
 figure, one can conclude that the these results essentially improve the good agreement with data
 in comparison with the CDP bounds at the wide range of $Q^{2}$
 values. We observe that this ratio is comparable with the CDP
 bounds at some values of $Q^{2}$ and  it is compatible with the experimental data in a wide range of
$Q^{2}$ values.\\
To emphasize the size of the CDP bounds we show that the ratio
$F_{L}/F_{2}$ and  $R$ have a maximum behavior when the proton
structure function has a power law behavior. In Figs.(3) and (4)
the ratio $F_{L}/F_{2}$ and $R$ obtained based on the gluon and
singlet exponents at $W=230~\mathrm{GeV}$. These behaviors  are in
good agreements with the CDP bounds when applying the uncertainty
principle at moderate $Q^{2}$ values. These results indicate a
decrease of the ratios for small $Q^{2}$ values which it is
require as to the electromagnetic gauge invariance. On the other
hand, at large $Q^{2}$ values the exponent method is not
consistent with the experimental data.\\
Therefore we study the ratio of structure functions with respect
to the effective exponents. In Figs.(5) and (6)  the ratio
$F_{L}/F_{2}$ and $R$ are plotted as related to the singlet
effective exponent $\lambda^{s}(Q^{2})$. It is seen that our
results based on the effective exponent at NNLO approximation,
over a wide range of $x$ and $Q^{2}$ values, are comparable with
the experimental data at low and moderate $Q^{2}$ values. At
high-$Q^{2}$ values, an overall shift between the HERA data and
the predictions is observed. This behavior can be resolved with an
adjustment of singlet exponent than one obtained with respect to
the effective and
phenomenological exponents via Eqs.(40) and (41) respectively.\\
The agreement between the method and the experimental data is good
until $Q^{2}=200~\mathrm{GeV}^{2}$. Because the singlet effective
exponents  (i.e., Eqs.(40) and (41)) parameterized only for
$Q^{2}< 200~\mathrm{GeV}^{2}$. One can conclude that exponent
defined for singlet distribution is larger than the gluon exponent
at large $Q^{2}$ values where this
behavior is not consistent with pQCD.\\
In figures (1) and (2) the behavior of the ratio $F_{L}/F_{2}$ and
$R$ at low $Q^{2}$ should eventually be tamed by the nonlinear
effects of the singlet density. The nonlinear corrections (NLCs)
to the ratio ${F_{L}}/{F_{2}}$ and $R$ are considered in a wide
range of $Q^{2}$ values in Fig.7. In this figure (i.e., Fig.7),
the effects of nonlinearity are investigated in the hot-spot point
($R=2~\mathrm{GeV}^{-1}$) in comparison with the linear behavior
from the $F_{2}$ parameterized. One can see that obtained
nonlinear corrections for these ratios are observable  at low
$Q^{2}$ values ($1<Q^{2}<10~\mathrm{GeV}^{2}$) and  comparable
with the H1 data in a wide range of
$Q^{2}$ values.\\
On the other hand, the nonlinear behaviors have been shown for the
ratio ${F_{L}}/{F_{2}}$ and $R$  with respect to the singlet
effective exponent $\lambda^{s}(Q^{2})$ in figure 8. The error
bands represent the uncertainty estimation coming from the $F_{2}$
parameterized. As one can see in this plot, the inclusion of the
nonlinear behavior by the effective exponent significantly change
the behavior of the ratio of ${F_{L}}/{F_{2}}$ and $R$ in a wide
range of $Q^{2}$ values. On can see an enhancement for the
moderate value of $Q^{2}$ and reduction for the small and large
values of $Q^{2}$. The results for the ratios clearly show
significant agreement with the HERA data over a wide range of $x$
and $Q^{2}$ variables. The comparison of the results reveals the
following conclusion: For a fixed value of the invariant mass $W$,
one see the same patterns for the ratios when the effective
exponent is $Q^{2}$-dependent. It is shown a pick around the
moderate value of $Q^{2}$,  $Q^{2} \sim 5~ \mathrm{GeV}^{2}$. One
of the main important results can be concluded from this figure is
the significant reduction in the ratio ${F_{L}}/{F_{2}}$ and $R$
at low $Q^{2}$ values caused by including the nonlinear effects
related to the singlet effective exponent in this analysis.


\subsection{6. Conclusion}

We presented the high-order corrections for ratio $F_{L}/F_{2}$
and $R$ with respect to the derivative of the proton structure
function into ${\ln}Q^{2}$. In this paper we have studied several
aspect of the proton structure function behavior at small $x$.
These ratios (i.e., $F_{L}/F_{2}$ and $R$) determined in the
kinematical region where $F_{2}$ has been parameterized. The
behavior of these ratios are in good agreement in comparison with
the experimental data  in a wide range of $Q^{2}$ values at a
fixed invariant mass. The value of these ratios are different from
the one obtained for the CDP bounds. Only at low $Q^{2}$ values
these results are comparable with the CDP bounds. Then we have
studied the effects of adding the nonlinear corrections to the
ratio $F_{L}/F_{2}$ and $R$ in this region. These results are
consistent with other experimental data, as we have discussed the
meaning of this finding from the point of
view of modified nonlinear behavior of these ratios.\\
Also a power-law behavior for the ratio of structure functions is
predicted. The results for a fixed exponent are comparable with
the CDP bounds  and HERA data at moderate $Q^{2}$ values. At low
and moderate $Q^{2}$ values an effective singlet exponent should
be considered. This analysis is also enriched with considering the
nonlinear contributions to the ratios related to the effective
singlet exponent. We have therefore solved the ratio $F_{L}/F_{2}$
and $R$ with the nonlinear shadowing term included in order to
determine the behavior of the effective singlet exponent at low
values of $Q^{2}$. These results show that the data can be
described in new colliders taking shadowing corrections to the
effective exponent into account. \\

\subsection{ACKNOWLEDGMENTS}
Authors are grateful the Razi University for financial support of
this project and also G.R.Boroun is thankful the CERN theory
department for their hospitality and support during the
preparation of this paper.\\

\subsection{Appendix A}
The kernels presented for the quark and gluon sectors, denoted by
$\Phi$ and $\Theta$ respectively at LO up to NNLO,
\begin{eqnarray}
\Theta_{qg}(x,Q^{2})&=&P_{qg}(x,\alpha_{s})\otimes
x^{\lambda_{g}},\nonumber\\
\Phi_{qq}(x,Q^{2})&=&P_{qq}(x,\alpha_{s})\otimes x^{\lambda_{s}},
\end{eqnarray}
have the following form at the leading order approximation  as:
\begin{eqnarray}
\Phi_{qq}(x,Q^{2})&=&\frac{\alpha_{s}}{4\pi}\{4+\frac{16}{3}\ln(\frac{1-x}{x})+\frac{16}{3}\int_{x}^{1}{\frac{z^{\lambda_{s}}-{z}^{-1}}{1-z}dz}\nonumber\\
&&-\frac{8}{3}\int_{x}^{1}{(1+z)z^{\lambda_{s}}dz\}},\nonumber\\
\Theta_{qg}(x,Q^{2})&=&\frac{\alpha_{s}}{4\pi}\frac{20}{9}\int_{x}^{1}{(z^2+(1-z)^2)z^{\lambda_{g}}dz}.
\end{eqnarray}
Also the longitudinal kernels at low-$x$ limit presented by the
following forms
\begin{eqnarray}
I_{L,q}(x,Q^{2})&=&\sum_{n=1}a(t)^{n}c_{L,q}^{n}(x)\otimes
x^{\lambda_{s}},\nonumber\\
I_{L,g}(x,Q^{2})&=&\sum_{n=1}a(t)^{n}c_{L,g}^{n}(x)\otimes
x^{\lambda_{g}}.
\end{eqnarray}
can be defined at  leading order approximation by:
\begin{eqnarray}
I_{L,q}(x,Q^{2})&=&\frac{\alpha_{s}}{4\pi}\int_{x}^{1}8n_{f}(1-z)z^{\lambda_{s}+1}dz,\nonumber\\
I_{L,g}(x,Q^{2})&=&\frac{\alpha_{s}}{4\pi}\int_{x}^{1}4C_{F}z^{\lambda_{g}+1}dz.
\end{eqnarray}
 \begin{table}[h]
\centering \caption{HERA data collected by two collaborations H1
and ZEUS. }\label{table:table1}
\begin{minipage}{\linewidth}
\renewcommand{\thefootnote}{\thempfootnote}
\centering
\begin{tabular}{|l|c|c|} \hline\noalign{\smallskip} HERA & $e^{+}p$ &  $e^{-}p$\\
\hline\noalign{\smallskip}
HERA I & 100 pb$^{-1}$ & 15 pb$^{-1}$   \\
HERA II& 150 pb$^{-1}$ & 235 pb$^{-1}$  \\

\hline\noalign{\smallskip}
\end{tabular}
\end{minipage}
\end{table}

\begin{table}[h]
\caption{ The effective Parameters at low $x$ for
$0.15~\mathrm{GeV}^{2}<Q^{2}<3000~\mathrm{GeV}^{2}$ provided by
the following values. The fixed  parameters are defined by the
Block-Halzen fit to the real photon-proton cross section as
$M^{2}=0.753 \pm 0.068~ \mathrm{GeV}^{2}$ and $\mu^2 = 2.82 \pm
0.290~ \mathrm{GeV}^{2}$.}
\begin{tabular} {cccc}
\toprule \\  \multicolumn{2}{c}{parameters \quad \quad \quad ~~~~~~~~~~~~~~~~value}    \\ &&&\\ \hline \\ &&&\\
$a_{00}$& \quad  $2.550\times 10^{-1}~\pm 1.60\times10^{-2}$ \\

$a_{01}$& \quad  $1.475\times 10^{-1}~\pm 3.025\times10^{-2}$\\&&&\\

  $a_{10} $  &   \quad  $8.205\times 10^{-4}~~  \pm  4.62\times10^{-4} $  \\

  $a_{11} $  &   \quad   $-5.148\times 10^{-2}\pm 8.19\times10^{-3}$  \\

  $a_{12}$   &    \quad  $-4.725\times 10^{-3}\pm 1.01\times10^{-3}$   \\  &&&\\

 $a_{20}$   &   \quad   $2.217\times 10^{-3}\pm 1.42\times10^{-4} $ \\

 $a_{21}$   &   \quad   $1.244\times 10^{-2}\pm 8.56\times10^{-4}$  \\

 $a_{22}$    &    \quad  $5.958\times 10^{-4}\pm 2.32\times10^{-4} $ \\ &&& \\

$n$& \quad  $11.49\pm 0.99$ & &\\

$\lambda$& \quad  $2.430~\pm 0.153$ & &\\

$\chi^{2}(\mathrm{goodness~ of~ fit})$ &  \quad  $0.95$ & &\\

\hline

\end{tabular}
\end{table}
\begin{table}[h]
\centering \caption{The QCD coupling and corresponding $\Lambda$
parameter for $n_{f}=4$ at LO, NLO [33, 43] and NNLO analysis
[42].}\label{table:table2}
\begin{minipage}{\linewidth}
\renewcommand{\thefootnote}{\thempfootnote}
\centering
\begin{tabular}{|l|c|c|} \hline\noalign{\smallskip}  & $ \alpha_{s}(M_{Z}^{2})$ &$
\Lambda_{QCD}(MeV)$  \\
\hline\noalign{\smallskip}
LO & 0.1166 & 136.8 \\
NLO & 0.1166 & 284 \\
NNLO & 0.1155 & 235 \\
\hline\noalign{\smallskip}
\end{tabular}
\end{minipage}
\end{table}
\newpage{
\section{References}
1. M.Klein,  arXiv [hep-ph]:1802.04317.\\
2. N.Armesto et al., Phys. Rev. D {\bf100}, 074022 (2019).\\
3. A. Abada et al., [FCC Collaborations], Eur.Phys.J.C{\bf 79}, 474(2019).\\
4. H.Abramowicz et al.,[H1 and ZEUS Collaborations],
Eur.Phys.J.C{\bf 75}, 580(2015).\\
5. V.Andreev et al. [H1 Collaboration], Eur.Phys.J.C{\bf74}, 2814
(2014).\\
6. A. H. Mueller and J. Qiu, Nucl. Phys. B\textbf{268}, 427(1986);
L. V. Gribov, E. M. Levin and M. G. Ryskin, Phys.
Rep.\textbf{100}, 1(1983).\\
7. G.R.Boroun and S.Zarrin, Eur.Phys.J.Plus {\bf128}, 119(2013);
G. R. Boroun and B. Rezaei, Chin. Phys. Lett.{\bf32},
111101(2015); B. Rezaei and G. R. Boroun, Phys. Lett. B{\bf692},
247 (2010);
 G. R. Boroun, Eur. Phys. J. A{\bf43}, 335(2010); G.R.Boroun, Eur.Phys.J. A{\bf42}, 251 (2009).\\
 8. M.Devee, arXiv[hep-ph]:1808.00899; M.Devee and J.K.sarma, Nucl.Phys. B{\bf885}, 571(2014); M.Lalung et al., Nucl.Phys. A{\bf984}, 29 (2019);
 P.Phukan et al., Nucl.Phys. A{\bf968}, 275 (2017).\\
 9. R.Wang and X.Chen, Chin.Phys. C{\bf41}, 053103 (2017); J.Lan et al., arXiv[nucl-th]:1907.01509.\\
10. N. N. Nikolaev and W. Sch$\ddot{a}$fer, Phys. Rev. D{\bf74}, 014023(2006).\\
11. M.Kuroda and D.Schildknecht, Phys.Lett. B{\bf618}, 84(2005); M.Kuroda and D.Schildknecht, Acta Phys.Polon. B{\bf37}, 835(2006).\\
12. M.Kuroda and D.Schildknecht, Phys.Lett. B{\bf670}, 129(2008); M.Kuroda and D.Schildknecht, Phys.Rev. D{\bf96}, 094013(2017).\\
13. D.Schildknecht and M.Tentyukov, arXiv[hep-ph]:0203028; M.Kuroda and D.Schildknecht, Phys.Rev. D{\bf85}, 094001(2012).\\
14. D.Schildknecht, Mod.Phys.Lett.A{\bf 29}, 1430028(2014).\\
15. M.Kuroda and D.Schildknecht, Int. J. Mod. Phys. A{\bf31}, 1650157 (2016).\\
16. Amir H.Rezaeian and I.Schmidt, Phys.Rev. D{\bf88}, 074016 (2013).\\
17. J.R.Forshaw et al., JHEP {\bf0611}, 025(2006).\\
18. C.Ewerz et al., Phys.lett.B\textbf{720}, 181(2013).\\
19. C.Ewerz and O.Nachtmann, Phys.Lett.B\textbf{648}, 279(2007).\\
20. M.Niedziela and M.Praszalowicz, Acta Phys.Polon. B{\bf46}, 2019(2015).\\
21. F.D. Aaron et al. [H1 Collaboration], phys.Lett.B\textbf{665},
139(2008); Eur.Phys.J.C\textbf{71},1579(2011).\\
22. C.Ewerz et al., Phys.Rev. D{\bf77}, 074022(2008).\\
23. H.Abromowicz et al. [ZEUS Collaboration],
Phys.Rev.D\textbf{9}, 072002(2014).\\
24. S.Moch, J.A.M.Vermaseren, A.Vogt, Phys.Lett.B \textbf{606},
123(2005).\\
25. W.L. van Neerven, A.Vogt, Phys.Lett.B \textbf{490}, 111(2000).\\
26. A.Vogt, S.Moch, J.A.M.Vermaseren, Nucl.Phys.B \textbf{691}, 129(2004).\\
27. Yu.L.Dokshitzer, Sov.Phys.JETP {\textbf{46}}, 641(1977);
G.Altarelli and G.Parisi, Nucl.Phys.B \textbf{126}, 298(1977);
V.N.Gribov and L.N.Lipatov, Sov.J.Nucl.Phys. \textbf{15},
438(1972).\\
28. V.S.Fadin, E.A.Kuraev and L.N.Lipatov, Phys.Lett.B
\textbf{60}, 50(1975); L.N.Lipatov, Sov.J.Nucl.Phys. \textbf{23},
338(1976);
I.I.Balitsky and L.N.Lipatov, Sov.J.Nucl.Phys. \textbf{28}, 822(1978).\\
29. G.R.Boroun and B.Rezaei, Nucl.Phys.A{\bf990}, 244(2019).\\
30.  B.Rezaei and G.R.Boroun, Eur.Phys.J.A{\bf55}, 66(2019).\\
31. S.Moch, J.A.M.Vermaseren, A.Vogt, Phys.Lett.B \textbf{606},
123(2005).\\
32. G.R.Boroun, Phys.Rev. C{\bf97}, 015206 (2018); G.R.Boroun and B.Rezaei, Eur.Phys.J. C{\bf72}, 2221(2012).\\
33. M. M. Block, L. Durand and P. Ha, Phys. Rev.D{\bf 89}, no. 9,
094027 (2014).\\
34. L.P.Kaptari et al., Phys.Rev.D {\bf 99}, 096019(2019).\\
35. F. D. Aaron et al. [H1 and ZEUS Collaborations], JHEP{\bf
1001}, 109(2010).\\
36.  A. Donnachie, P.V. Landshoff, Phys. Lett. B {\bf550}, 160
(2002); B.Rezaei and G.R.Boroun , Int.J.Theor.Phys. {\bf57}, 2309(2018).\\
37. Martin M. Block et al., Phys. Rev. D {\bf84} , 094010
(2011).\\
38. M. Praszalowicz, Phys. Rev. Lett. {\bf106}, 142002(2011).\\
39. M. Praszalowicz, T. Stebel, JHEP{\bf 03}, 090(2013).\\
40. H1 and ZEUS Collaboration (F.D. Aaron et al.), JHEP{\bf 01},
109 (2010); H1 and ZEUS Collaboration (F.D. Aaron et al.), Eur.
Phys. J. C {\bf63}, 625(2009); H1 and ZEUS Collaboration (F.D.
Aaron et
al.), Eur. Phys. J. C {\bf64}, 561(2009).\\
41. K. J. Eskola et al., Nucl. Phys. B{\bf660}, 211(2003);  R.
Fiore, P. V. Sasorov and V. R. Zoller, JETP Letters {\bf96},
687(2013); R. Fiore, N. N. Nikolaev and V. R. Zoller,
JETP Letters {\bf99}, 363(2014).\\
42. A.D.Martin et al., Phys.Letts.B{\bf 604}, 61(2004).\\
43. S. Chekanov et al. [ZEUS Collaboration], Eur. Phys. J.C {\bf 21}, 443 (2001).\\
44. K Golec-Biernat and A.M.Stasto, Phys.Rev.D {\bf80},
014006(2009).\\
45. G.R.Boroun, Eur.Phys.J.Plus {\bf135}, 68(2020); G.R.Boroun and B.Rezaei, Eur.Phys.J. C{\bf73}, 2412(2013); Phys.Atom.Nucl.{\bf71}, 1077(2008); EPL{\bf100}, 41001(2012).\\
46. A. Y. Illarionov, A. V. Kotikov and G. Parente Bermudez, Phys.
Part. Nucl. {\bf39}, 307 (2008).\\
47.  V.Tvaskis et al., Phys.Rev.C{\bf 97}, 045204(2018);
Phys.Rev.Lett.{\bf98}, 142301(2007).\\

}
\begin{figure}
\includegraphics[width=0.55\textwidth]{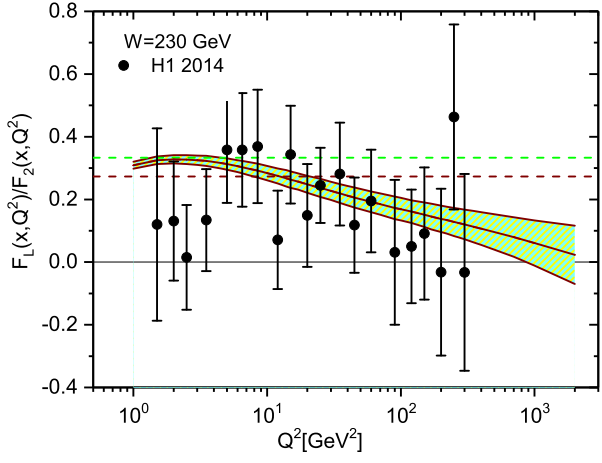}
\caption{The ratio $F_{L}/F_{2}$ extracted at  NNLO approximation
in comparison with the H1 data [30] as accompanied with total
errors. The results are presented  at fixed value of the invariant
mass W in the interval $1~\mathrm{GeV}^{2 }\leq Q^{2} <
3000~\mathrm{GeV}^{2}$ at low values of $x$. The shaded are
correspondent to uncertainties of the $F_{2}$ parameterization
(i.e., Table II). The dipole upper bounds  represented by the
dashed lines related to $\rho=1$ and $\frac{4}{3}$ respectively.
}\label{Fig1}
\end{figure}
\begin{figure}
\includegraphics[width=0.55\textwidth]{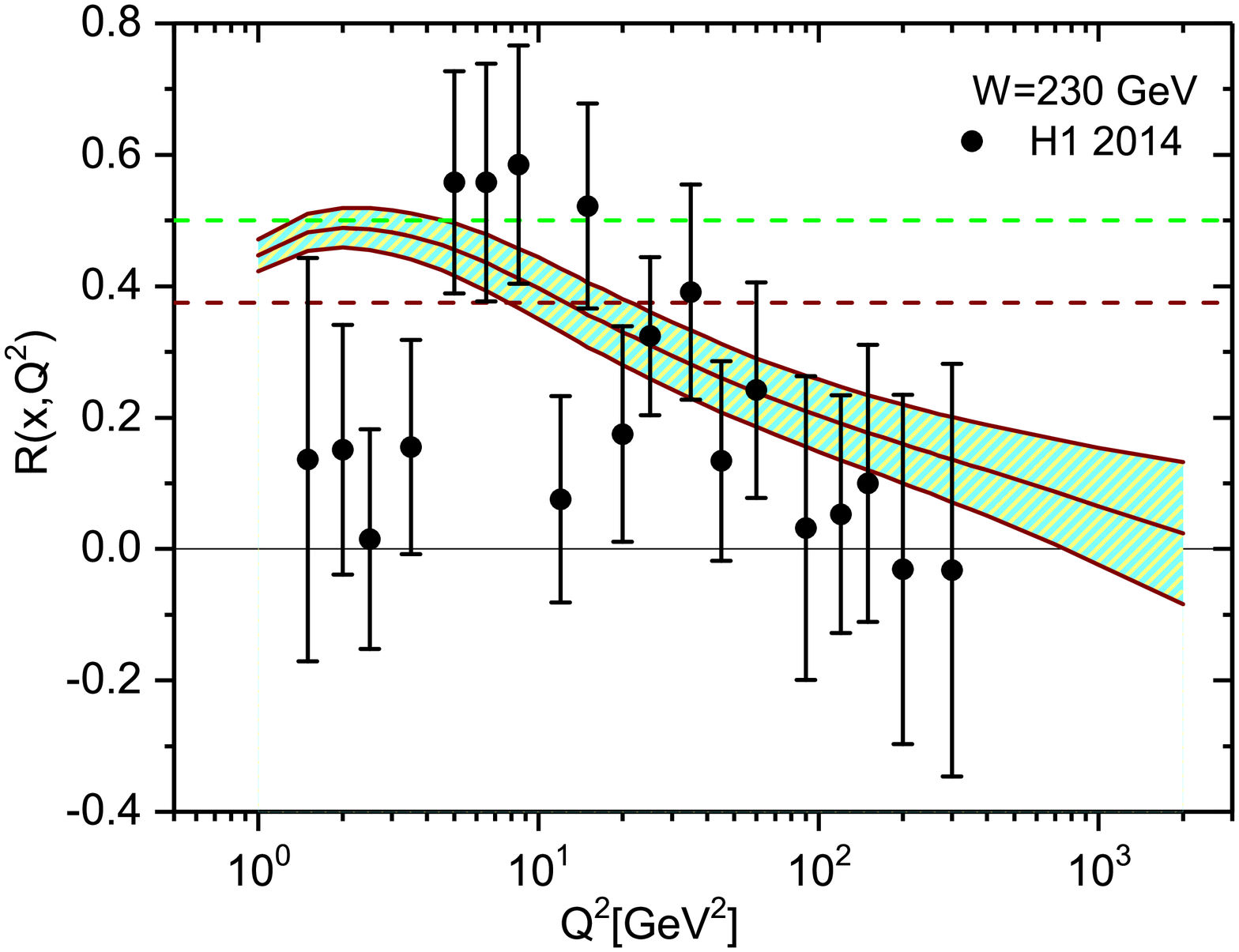}
\caption{The same as Fig.1 for the ratio $R$.}\label{Fig2}
\end{figure}
\begin{figure}
\includegraphics[width=0.55\textwidth]{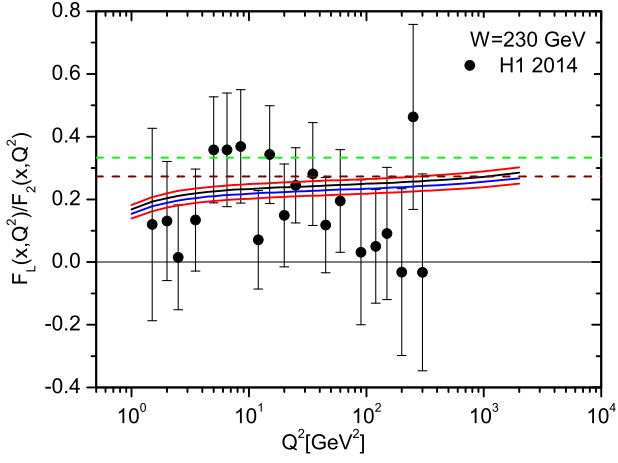}
\caption{The ratio $F_{L}/F_{2}$ extracted at  NNLO approximation
in comparison with the H1 data [30] as accompanied with total
errors. The results are presented  at fixed value of the invariant
mass W in the interval $1~\mathrm{GeV}^{2 }\leq Q^{2} <
3000~\mathrm{GeV}^{2}$ at low values of $x$. The solid lines are
defined with the fixed exponents. The singlet exponents
 are a dynamical quantity of the order of
$\lambda=0.27, 29, 31$ and $0.33$ from lower to upper curves
respectively. The dipole upper bounds represented by the dashed
lines related to $\rho=1$ and $\frac{4}{3}$ respectively.
}\label{Fig3}
\end{figure}
\begin{figure}
\includegraphics[width=0.55\textwidth]{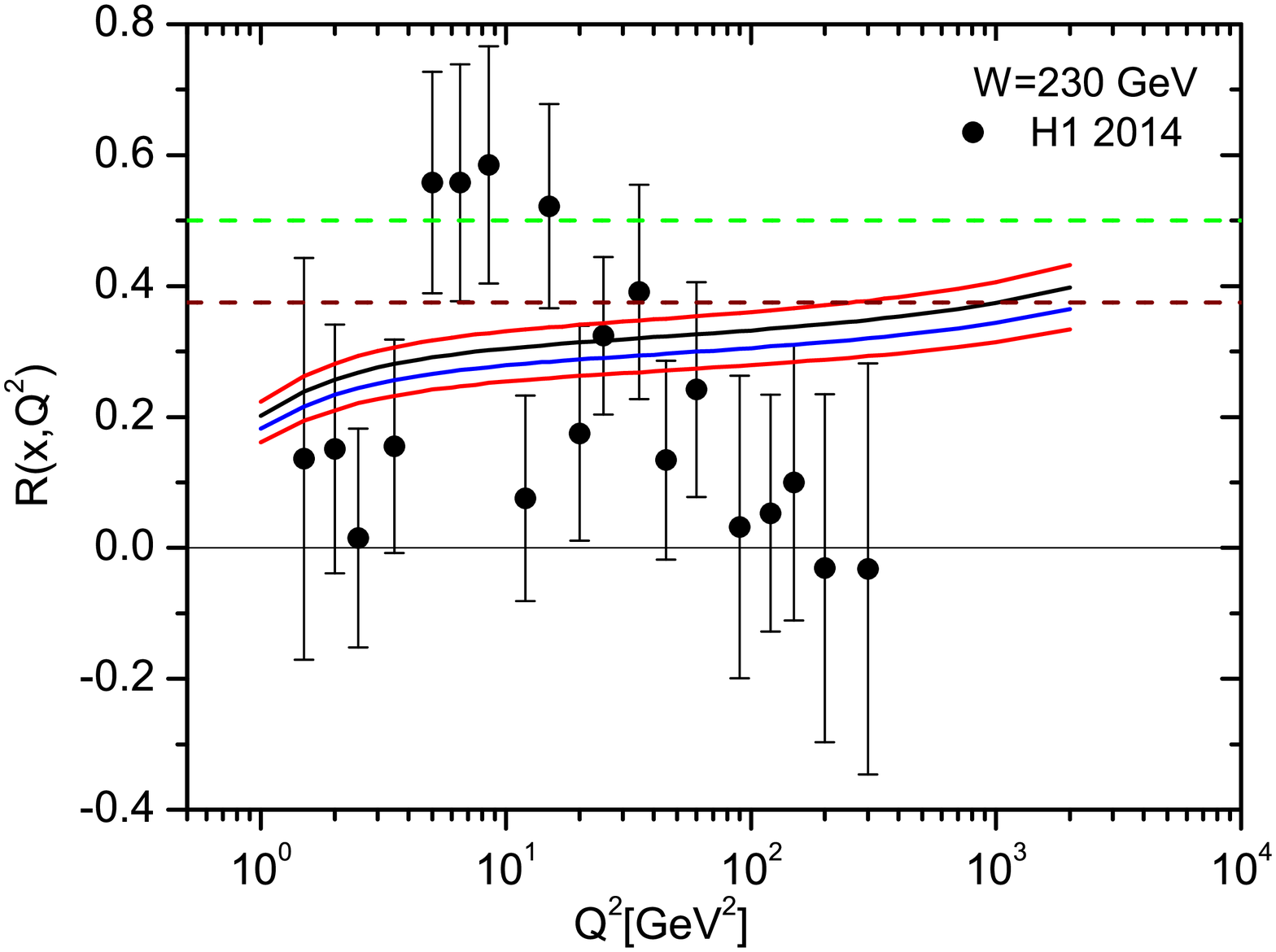}
\caption{The same as Fig.3 for the ratio $R$.}\label{Fig4}
\end{figure}
\begin{figure}
\includegraphics[width=0.55\textwidth]{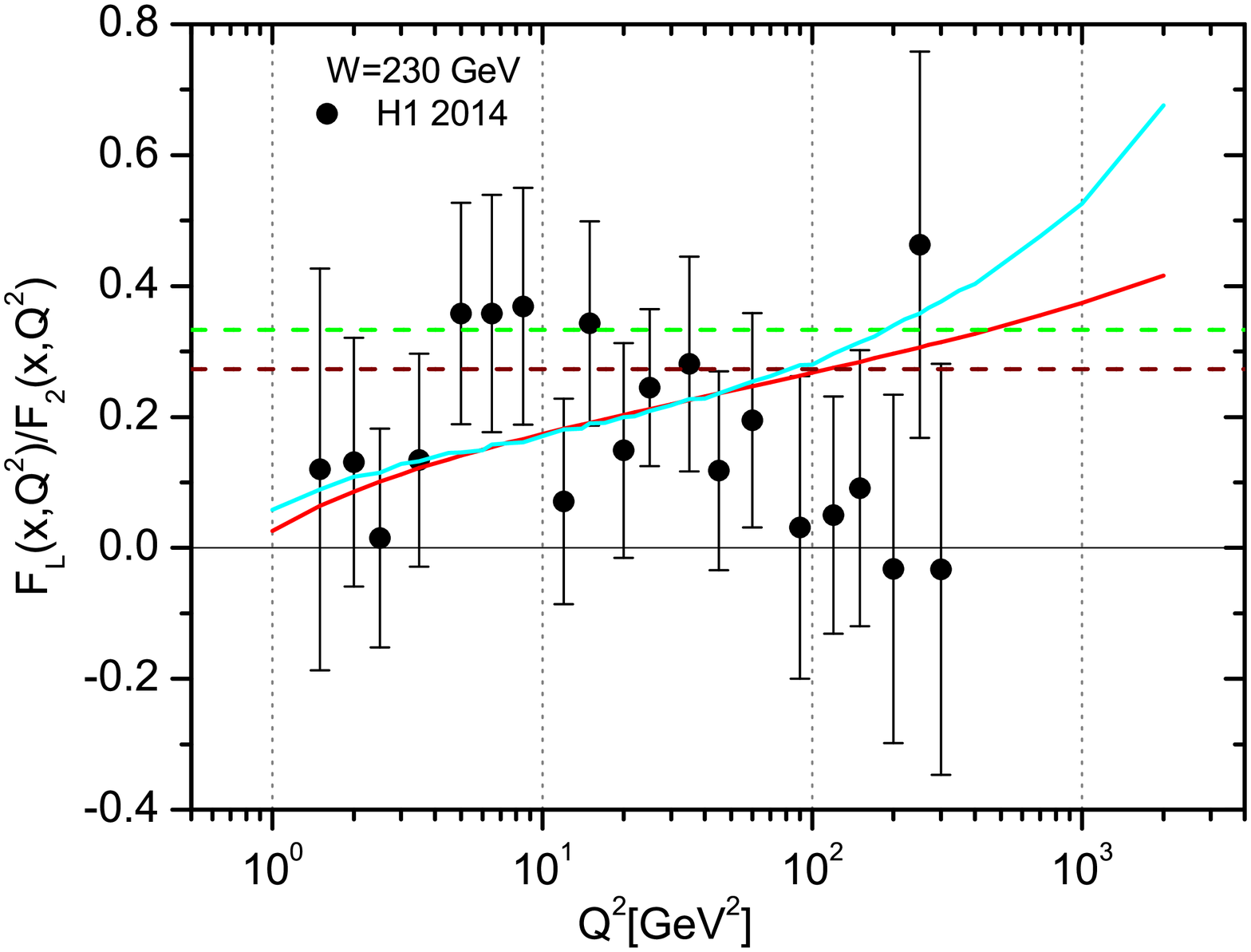}
\caption{The ratio $F_{L}/F_{2}$ extracted at  NNLO approximation
in comparison with the H1 data [30] as accompanied with total
errors. The results are presented  at fixed value of the invariant
mass W in the interval $1~\mathrm{GeV}^{2 }\leq Q^{2} <
3000~\mathrm{GeV}^{2}$ at low values of $x$. The solid lines are
defined with respect to the effective singlet exponents. The
exponents $\lambda^{s}(Q^{2})$ are parameterized as to Eqs.(40)
and (41) from lower to upper curves respectively. The dipole upper
bounds  represented by the dashed lines related to $\rho=1$ and
$\frac{4}{3}$ respectively.}\label{Fig5}
\end{figure}
\begin{figure}
\includegraphics[width=0.55\textwidth]{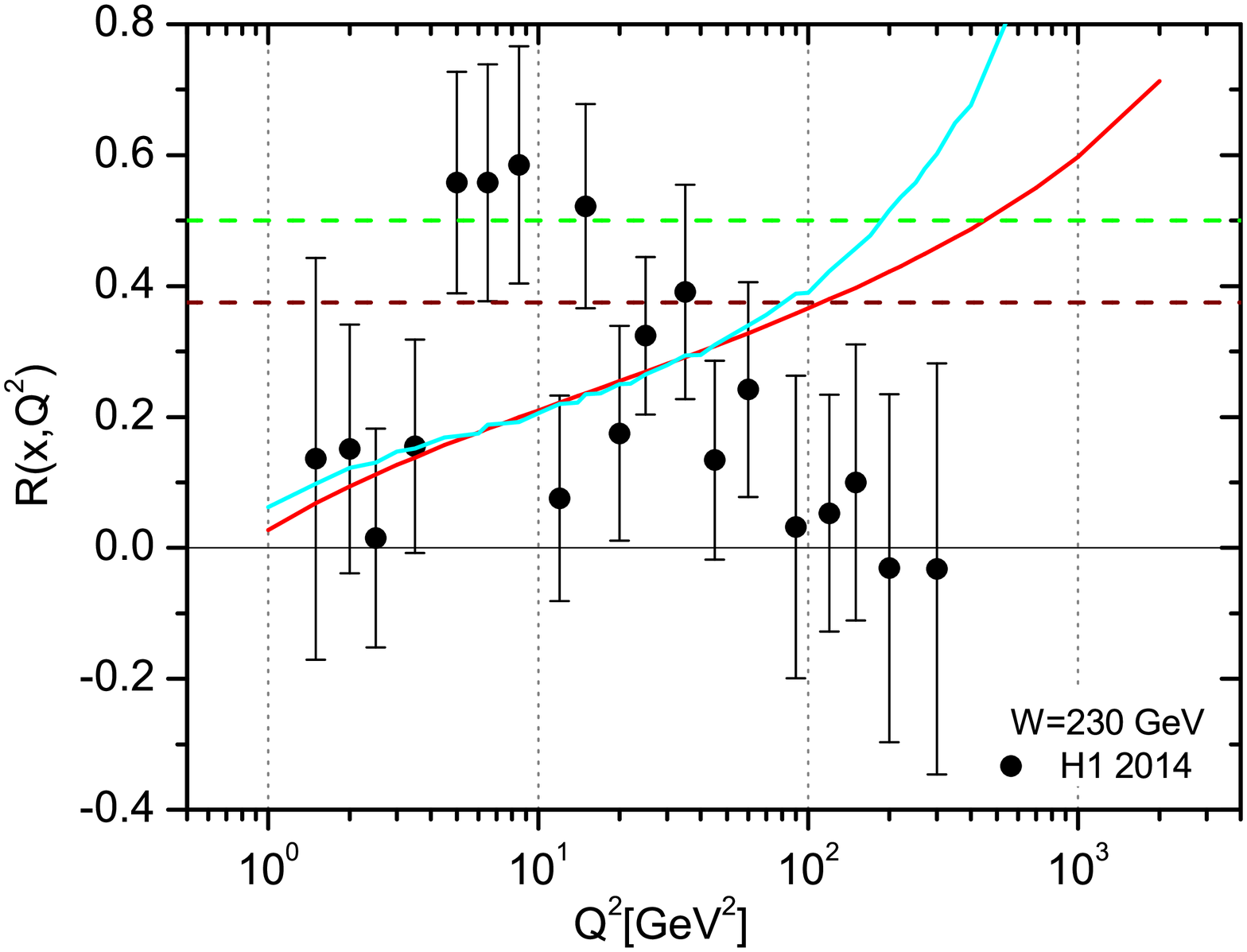}
\caption{The same as Fig.5 for the ratio $R$.}\label{Fig6}
\end{figure}
\begin{figure}
\includegraphics[width=0.56\textwidth]{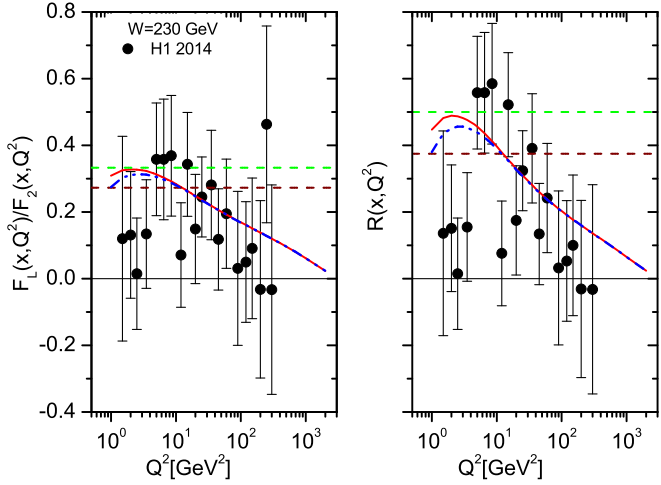}
\caption{The nonlinear corrections to the ratio of structure
functions (left, $F_{L}(x,Q^{2})/F_{2}(x,Q^{2})$) and (right,
$R(x,Q^{2})$) extracted at NNLO approximation in comparison with
the H1 data [16] as accompanied with total errors. The results are
presented at fixed value of the invariant mass W in the interval
$1~\mathrm{GeV}^{2 }\leq Q^{2} < 3000~\mathrm{GeV}^{2}$ at low
values of $x$. The dash-dot line is defined with the nonlinear
corrections in comparison with the linear (i.e., solid line). The
dipole upper bounds represented by the dashed lines related to
$\rho=1$ and $\frac{4}{3}$ respectively.}\label{Fig7}
\end{figure}
\begin{figure}
\includegraphics[width=0.56\textwidth]{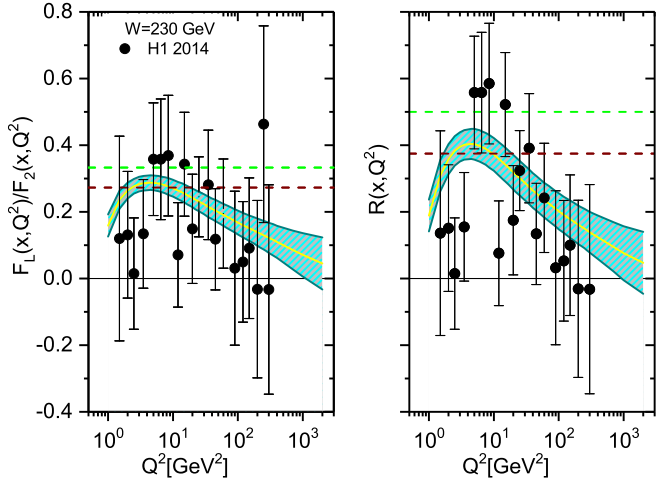}
\caption{The nonlinear corrections to the ratio of structure
functions (left, $F_{L}(x,Q^{2})/F_{2}(x,Q^{2})$) and (right,
$R(x,Q^{2})$) with respect to the effective exponent
$\lambda^{s}(Q^{2})$ [39] extracted at NNLO approximation in
comparison with the H1 data [30] as accompanied with total errors.
The results are presented at fixed value of the invariant mass W
in the interval $1~\mathrm{GeV}^{2 }\leq Q^{2} <
3000~\mathrm{GeV}^{2}$ at low values of $x$. The error bands
represent the uncertainty estimation coming from the $F_{2}$
parameterization. The dipole upper bounds represented by the
dashed lines related to $\rho=1$ and $\frac{4}{3}$
respectively.}\label{Fig8}
\end{figure}

\end{document}